\documentclass[twocolumn,preprintnumbers,floatfix,amsmath,a4paper,amssymb,nofootinbib, showpacs]{revtex4}

\usepackage{amsmath}
\usepackage{graphicx}
\usepackage{dcolumn}
\usepackage{bm}
\usepackage{amssymb}
\usepackage{epstopdf}
\usepackage{color}
\usepackage{courier}

\begin{document}

\title{Sample space reducing cascading processes produce the full spectrum of scaling exponents}{}

\author{Bernat Corominas-Murtra$^{1,2}$, Rudolf Hanel$^{1,2}$ and Stefan Thurner$^{1,2,3,4}$
}
\email{stefan.thurner@meduniwien.ac.at} 

\affiliation{
$^1$ Section for the Science of Complex Systems, CeMSIIS, Medical University of Vienna,
Spitalgasse 23, A-1090, Vienna, Austria\\
$^2$ Complexity Science Hub Vienna, Josefst\"adterstrasse 39, 1080 Vienna, Austria
$^3$ Santa Fe Institute, 1399 Hyde Park Road, Santa Fe, NM 87501, USA\\
$^4$ IIASA, Schlossplatz 1, 2361 Laxenburg, Austria\\
}

\date{Version \today}

\begin{abstract}
Sample Space Reducing (SSR)  processes are simple stochastic processes that offer a new route to understand scaling in path-dependent processes.
Here we define a cascading process that generalises the recently defined SSR processes and is able to produce power laws with arbitrary exponents. We demonstrate analytically that the frequency distributions of states are power laws with exponents that coincide with the multiplication parameter of the cascading process. In addition, we show that imposing energy conservation in SSR cascades allows us to recover Fermi's classic result on the energy spectrum of cosmic rays, with the universal exponent $-2$, which is independent of the multiplication parameter of the cascade. Applications of the proposed process include fragmentation processes or directed cascading diffusion on networks, such as rumour or epidemic spreading.
\end{abstract}

\pacs{ 89.75.Da, 
05.40.-a, 		 
05.10.Ln, 		 
05.40.Fb 		 
}
\maketitle

\section{Introduction}
Practically all complex adaptive systems exhibit fat-tailed distribution functions in the statistics of their dynamical variables.
Often these distribution functions are exact or almost exact asymptotic power-laws, $p(x) \sim x ^{-\lambda}$. 
While Gaussian statistics can often be traced back to a single origin, the central limit theorem, for the origin of power laws there exist 
several routes. These include
\begin{figure}[ht!]
\includegraphics[width= 8.7cm]{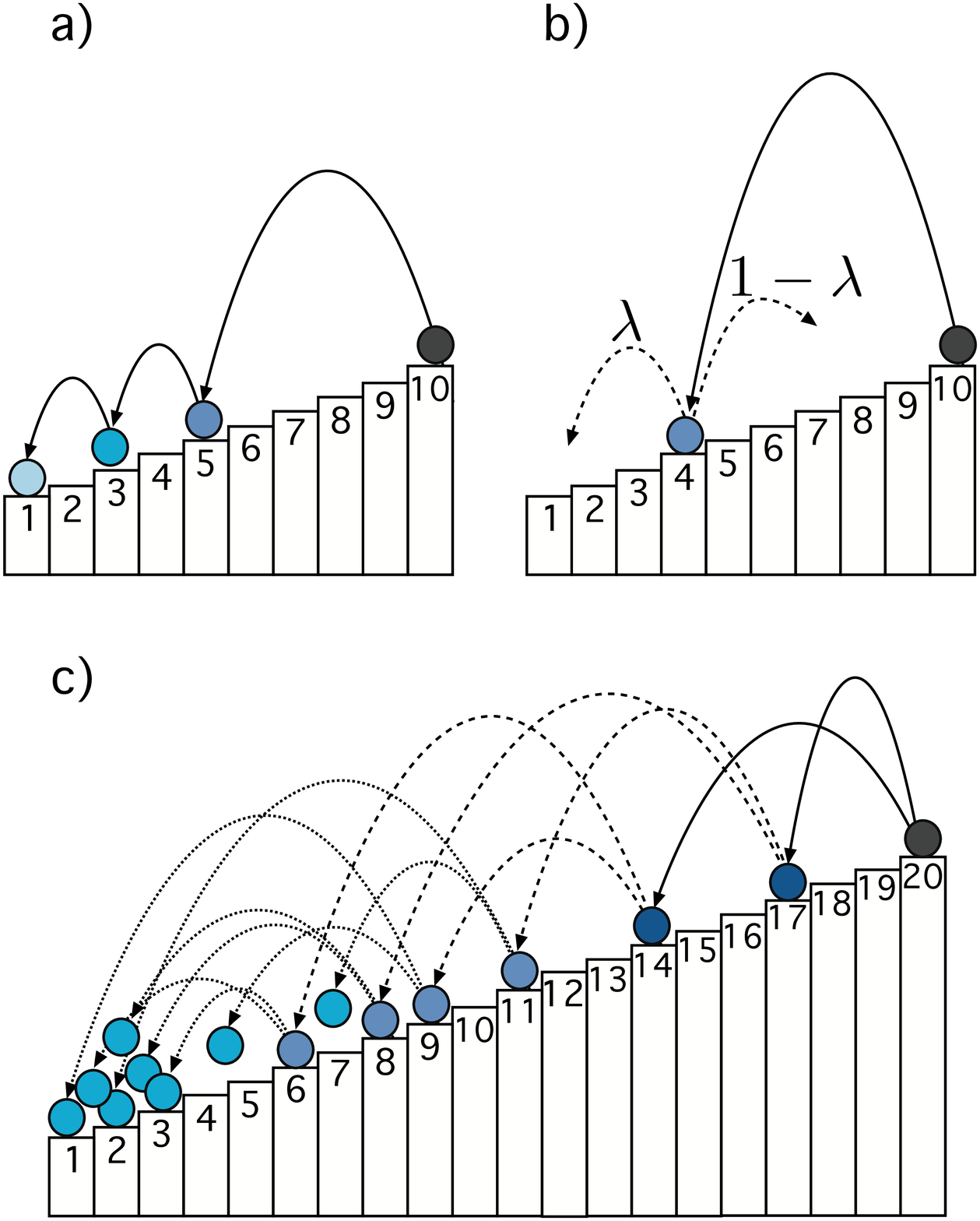}
\caption{(a) SSR process. A ball starts at the highest state $i = N = 10$, and may sequentially jump randomly toward
any lower state. Once the ball hits the lowest state
$i = 1$ the process starts again at state $N$. If repeated many times, the probability of visiting states is an exact
Zipf's law, $p(i) \propto i^{-1}$. (b) If with probability $1-\lambda$, $\lambda\in  [0, 1])$ the ball is allowed to jump to any state, 
the visiting probability becomes $p(i)\propto i^{-\lambda}$. $1-\lambda$ can be seen as the noise strength in a noisy SSR process, 
see \cite{Corominas-Murtra:2015}
(c) SSR cascading process with a multiplicative parameter $\mu=2$. A ball starts at the highest state, $i=N=20$, and splits into $\mu$ new 
balls which independently jump to lower states as before. Whenever a ball hits a state it creates $\mu$ balls which  continue their random 
jumps. It becomes a cascading or an avalanche process, and the visiting probability scales as $p(i)\propto i^{-\mu}$. We observe that (b) is automatically recovered if the multiplication parameter is $\mu<1$.
}\label{fig:AvalanchesDiagram}
\end{figure}
%
(i) Yule-Simon processes (preferential attachment) \cite{Yule:1925, Simon:1955, Barabasi:1999},
(ii) multiplicative processes with constraints, see \cite{Mietzenmacher:2003, Newman:2005},
(iii) criticality \cite{Stanley:1987},
(iv) self-organized criticality and cascading processes \cite{Bak:1987,Kadanoff:1989,Jensen:1996,Christiansen:2005}, 
(v) constraint optimization \cite{Mandelbrot:1953,Topsoe:2001,Corominas-Murtra:2011},
and 
(vi) sample space reducing (SSR) processes \cite{Hanel:2014,Corominas-Murtra:2015, Corominas-Murtra:2016}.
Most of these mechanisms are able to explain specific values of the exponent $\lambda$ or a range of exponents.
None of them however explain the full range of exponents from zero to infinity, $\lambda\in[0, \infty)$ in a straightforward way. 
Depending on the generative process exponents belong to different ranges. For example, exponents near critical points are 
often fractional, and within the range  $\lambda\in[1/2,5/2]$ \cite{Stanley:1987}. Many exponents for avalanche processes are 
found within a  range of $\lambda\in(0,3)$ \cite{Maya:1996}.  Some processes, like the preferential attachment, can formally 
explain  a wide range of exponents, although deviations from the standard values $\lambda\in(2,3.5)$ are hard to map to realistic underlying 
stochastic dynamics \cite{Jackson:2010}.
Here we show that the combination of cascading processes with SSR processes is able to do exactly that, 
to provide a single one-parameter model that produces the full spectrum of all possible scaling exponents.
The parameter is nothing but the multiplication ratio of the cascading processes. Finally, we show that generic disintegration processes can be mapped one-to-one to SSR cascades with an over imposed condition of conservation of whatever magnitude is represented by the states. This mapping allows us to derive a remarkable result: The histogram of visits to each state follows an exponent $-2$, regardless the multiplication parameter. This result may have important consequences in order to understand generic properties of disintegration processes and the ubiquity of the exponent $-2$ in nature.This, for example,
 allows us to recover Fermi's classic result on the energy spectrum of cosmic rays, only appealing to combinatorial properties of the cascade.  In addition, we provide a rigorous proof of that result in the appendix A, as a new contribution to the study of the random partition of the interval. 

Cascading processes 
have played an important role in the understanding of 
power-law statistics in granular media \cite{Bak:1987,Kadanoff:1989, Jensen:1996, Christiansen:2005, Frette:1996}, 
earth quakes \cite{Sornette:1989, Turcotte:1997, Corral:2004}, 
precipitation \cite{Peters:2006, Corral:2010}, 
dynamics  of combinatorial evolution \cite{Thurner:2010}, 
or failure in networks \cite{Boss:2004, Buldryev:2010, Thurner:2012}. 
The scaling exponents of the probability distribution functions of quantities  such as avalanche sizes, 
energy distributions, visiting times, event durations, etc., are found within a relatively narrow band. 
Cascading processes are often history-dependent processes in the sense that for a particular event taking place the temporal order of microscopic 
events is important. Recent progress in the understanding of the  generic statistics of history-dependent processes 
and their relation to power laws was made in \cite{Hanel:2013,Hanel:2014,Corominas-Murtra:2015, Corominas-Murtra:2016}. 
Maybe the simplest history-dependent processes are the sample space reducing (SSR) processes 
\cite{Corominas-Murtra:2015}, which explain the origin of scaling in a very simple and intuitive way. They have been used in 
various  applications in computational linguistics \cite{Thurner:2016}, fragmentation processes \cite{Corominas-Murtra:2015}, 
and diffusion on directed networks and search processes  \cite{Corominas-Murtra:2016}.  

\section{SSR processes and cascades}
A SSR process is a stochastic processes whose sample space reduces as it evolves in time. 
They can be depicted in a simple way, see figure  \ref{fig:AvalanchesDiagram}(a). 
Imagine a set of $N$ states in a system, labelled by $i=1,2,\cdots,N$. The states are ordered by the label. 
The only rule that defines the SSR process is that transitions between states may only occur from higher to lower labels. This means 
that transition from state $j\to i$ is possible only if label $j>i$. When the lowest state $i=1$ is reached, the process 
stops or is re-started. 
It was shown in \cite{Corominas-Murtra:2015} that this dynamics leads to a Zipf's law in the frequency of state visits, i.e. the probability to visit state 
$i$ is given by $p(i) = i^{-1}$.
This scaling law is extremely robust and occurs for a wide class of prior probabilities \cite{Corominas-Murtra:2016}. 
We now show that the combination of SSR processes with the simplest cascading process allows us to obtain a mechanism that 
can produce power laws with any scaling exponent. We will comment on how the model can be used to 
recover the Fermi's classic result on the cosmic ray spectrum \cite{Fermi:1949, Longair:2008}. This is possible 
by imposing energy conservation on SSR cascading processes. We discuss other potential cases where the theory of SSR cascades might apply. 

\subsection{SSR cascades}
\label{Sec:Cascades}
To define SSR cascades, imagine a system with a set of $N$ ordered states, each of which has a {\em prior} probability of appearing, $q_1, . . .,q_{N-1}$, being state $N$ the starting point of the cascade. 
The process starts at $t=0$ with $\mu$ balls at state $N$ jumping to any state $i_1,. . .,i_\mu<N$ with a probability proportional to $q_i, . . .,q_\mu$, respectively. 
Suppose that the $\mu$ balls landed on states $i_1, . . i_\mu$, respectively. At the next timestep, $t=1$,  each of these $\mu$ balls divide into $\mu$ new balls which all jump  to any state {\em below} their original state. The multiplicative process continues downwards. Whenever a ball hits the lowest state, it is eliminated from the system. 
Effectively we superimpose a multiplicative process that is characterized by the multiplicative parameter $\mu$, and the SSR process described above, see figure  \ref{fig:AvalanchesDiagram}(c).
The case $\mu=1$ is exactly the standard SSR process, where no new elements are created, and the case $\mu<1$ corresponds to the {\em noisy} SSR, where there is the possibility that the process gets cut at some step, see figure \ref{fig:AvalanchesDiagram}(b).

\begin{figure}[ht!]
\includegraphics[width= 8.0cm]{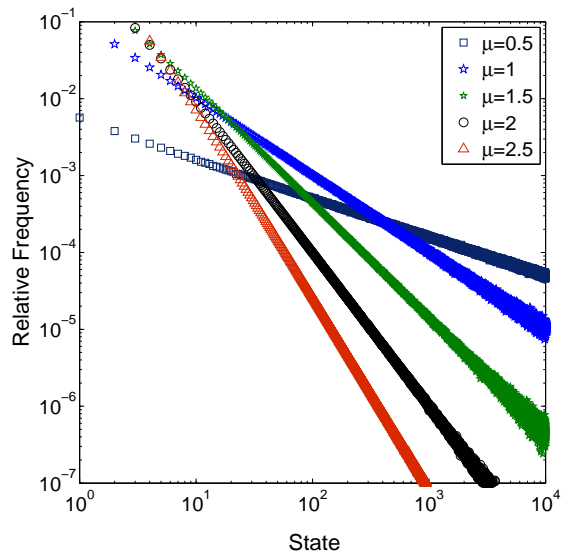}
\caption{Normalized histograms of state visits of $N=10,000$ states, obtained from numerical simulations of SSR cascades with a multiplicative 
parameters $\mu=0.5$ (dark blue squares), $\mu=1$ (blue circles), $\mu=1.5$ (green circles), $\mu=2$ (black diamonds) and $\mu=2.5$ (red triangles). 
Histograms are clearly power laws, $\sim i^{-\mu}$. 
Curves were fitted using the matlab \texttt{rplfit} package using likelihood estimations for all exponents \cite{Hanel:2016}. 
}
\label{fig:AvalanchesPlots}
\end{figure}

The derivation of the visiting distribution of this cascading SSR process follows the arguments found in \cite{Corominas-Murtra:2016}. 
We first define the cumulative prior distribution function $g(k)$, 
\begin{eqnarray}
g(k)=\sum_{i\leq k}q_i\quad. \nonumber
\end{eqnarray}
Without a multiplication factor $\mu$ the transition probabilities $p(i|j)$ determine the probability to reach state $i$ at timestep $t+1$, given 
that the system is in state $j$ at time $t$, and are given by
\begin{equation}
p(i|j)=\left\{
\begin{array}{cl}
	\frac{q_i}{g(j-1)}& {\rm for}\; i<j \\
	0 			& {\rm for}\; i \geq j \quad.
\end{array}
\right.
\label{eq:P(a_k|a_i)}
\end{equation}
In a SSR cascade, if there is an element sitting at state $j$ at time $t$, there are now $\mu$ trials to reach any state $i<j$ at $t+1$. 
Since the number of particles is not conserved throughout the process, we talk about 
the {\em expected} number of jumps from $j$ to $i$. Since the jumps from $j$ to $i$ of each ball is independent, the expected number of jumps from $j$ to $i$ we denoted by $n(j\to i)$ can be approximated as follows: 
\begin{equation}
	n(j\to i) =	\mu p(i|j)  \quad.
\label{eq:P(a_k|a_i)eta}
\end{equation}
We denote  the expected number of elements that will hit state $i$ in a given SSR cascade by $n_i$. 
Up to a factor the sequence $n_1, . . .,n_{N}$ is identical to the histogram of visits. 
From equations (\ref{eq:P(a_k|a_i)}) and (\ref{eq:P(a_k|a_i)eta}) we get 
\begin{equation}
	n_i=\sum_{j>i}n(j\to i)n_j  = \mu q_i\sum_{j>i}\frac{n_j}{g(j-1)}   \quad . \nonumber
\end{equation}
By subtracting $n_{i+1}-n_i$ and after re-arranging terms we find
\begin{equation}
	\frac{n_{i+1}}{q_{i+1}}\left(1+\mu\frac{q_{i+1}}{g(i)}\right)=\frac{n_i}{q_i} \quad ,\nonumber
\end{equation}
or, when applied iteratively 
\begin{equation}
	n_i=n_1\frac{q_i}{q_1}\prod_{1<j\leq i}\left(1+\mu\frac{q_j}{g(j-1)}\right)^{-1}\,.\nonumber
\end{equation}
Since  $\mu\frac{q_j}{g(j-1)}$ is typically small the product term is well approximated by
\begin{eqnarray}
	\prod_{1<j\leq i}\left(\cdots\right)^{-1}
	&=&\exp\left[-\sum_{1<j\leq i}\log\left(1+\mu\frac{q_j}{g(j-1)}\right)\right]  \nonumber\\
	&\approx& \exp\left(- \mu \sum_{1<j\leq i} \frac{q_j}{g(j-1)}\right)  \nonumber\\
	&\approx& \exp\left(- \mu \int_{1}^i \frac{dg}{dx}\frac{1}{g(x)}dx \right)  \nonumber\\
	&\approx& \exp\left(-\mu\log \frac{g(i)}{q_1} \right)  \nonumber\\
	&=&\left(\frac{g(i)}{q_1}\right)^{-\mu}\quad ,\nonumber
\end{eqnarray}
where we used $q_j \sim  \left. dg/dx\right|_{j}$ and $\log(1+x) \sim  x$.
Finally, we have 
\begin{equation}
	n_i \sim \frac{n_1}{q_1^{1-\mu}}\left(\frac{q_i}{g(i)^\mu}\right)\quad.
\label{eq:General_Noise}
\end{equation}
For equal prior probabilities, $q_i=\frac{1}{N-1}$ for all states, we get the expected visiting frequency to be
\begin{equation}
n_i\sim i^{-\mu}\quad.\nonumber
\end{equation}
The multiplication factor $\mu$ becomes the scaling exponent, for $\mu=1$ the standard SSR processes is recovered  \cite{Corominas-Murtra:2015}. 
Figure (\ref{fig:AvalanchesPlots}) shows numerical results which are in perfect agreement with the theoretical predictions.   
Note that the argument also holds for non-integer $\mu$, where, on average, $\mu$ balls are created at every step. In the numerical implementation, a non-integer $\mu$ is introduced as follows: Let $\mu=\lfloor \mu\rfloor+\delta$, with $\delta<1$. Then, with probability $\delta$,  $\lfloor \mu\rfloor+1$ balls are created and, with probability $1-\delta$, $\lfloor \mu\rfloor$ balls are created.
Also the case of multiplication factors $\mu<1$ are possible, reproducing the previously defined noisy SSR case, see figure (\ref{fig:AvalanchesDiagram}b) and \cite{Corominas-Murtra:2015}. In this situation at each step the process can be restarted with the  probability $1-\mu$.

\begin{figure}
\includegraphics[width= 8.0cm]{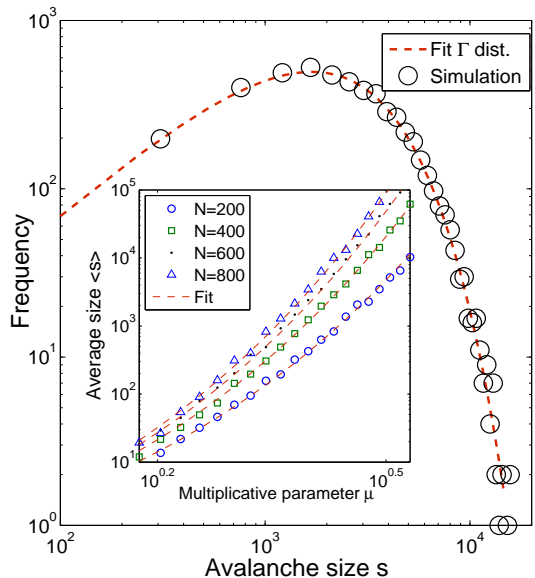}
\caption{Cascade size distribution after $3,000$ realisations of individual SSR cascades with $\mu=2.5$ and $N=10^4$ states (circles), with a 
fit based on the $\Gamma$ distribution $f(s)$ (red dashed line) given in equation  (\ref{eq:GammaDist}). 
Inset: dependence of $\langle s \rangle$ on $\mu$ (from 1.5 to 3.5) for four different system sizes, $N=200$ (blue circles), $N=400$ (green circles), $N=600$ (black diamonds) and $N=800$ (blue triangles). Dashed curves are fits of the average size $\langle s\rangle$ given in equation (\ref{eq:GammaDist}).}
\label{fig:AvalancheSize}
\end{figure}

We numerically compute the cascade size distribution as a function of the  number of states $N$ and $\mu$.  
For a given realisation of a cascade $\psi$ with initial sample space $N$ and multiplicative parameter $\mu$, starting with a single element at $N$,
we define the {\em cascade size}, $s_{\mu, N}(\psi)$, as {\em the number of elements of the cascade $\psi$ that reach state} $1$, $n_1^{(\psi)}$. 
Numerical analysis suggests that the cascade size distribution $f(s_{m,\mu})$ can be well approximated by a  $\Gamma$ distribution \cite{Feller:1959}.
 For the sake of simplicity, we drop the subscripts  
$\mu$ and $N$ for $s$.  
We thus find a purely phenomenological equation that reads
\begin{equation}
f(s)\propto s^{\alpha-1}e^{-\lambda s},\quad\langle s \rangle\propto  \frac{N^{\mu}}{e^{a\mu}},\quad \sigma^2 \propto \frac{N^{b\mu}}{e^{\left(\frac{1}{2}+a\right)\mu}} \langle s\rangle\;,
\label{eq:GammaDist}
\end{equation}
with  $a=0.82$, $b=0.9$, $\alpha=\langle s \rangle^2/\sigma^2$, $\lambda=\langle s\rangle/\sigma^2$.
Numerical results and fits are shown in figure  (\ref{fig:AvalancheSize}). The inset shows that the approximation for $\langle s \rangle$ is highly accurate.

\subsection{Energy conservation and the prevalence of $-2$ exponent}

\begin{figure}
\includegraphics[width= 8.3cm]{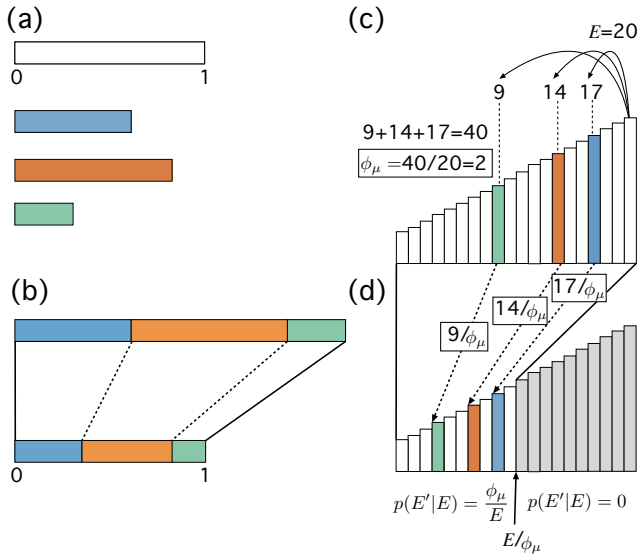}
\caption{SSR cascades with energy conservation and the random partition of the interval. (a) Consider a stick of length $1$ and chose three sticks (blue, orange and green) of random sizes between $0$ and $1$. (b) Put them together sequentially and glue them. This will create another stick made of the blue, orange and green sticks of size larger than $1$. Divide the stick by its size and one gets a stick of length $1$ partitioned in different random segments proportional to the sizes of the sticks chosen in (a). This process of random partition of the interval is exactly analogous to a SSR cascade with energy conservation. (c) At a given energy level ($20$) three balls jump downwards to a randomly chosen sites  ($17, 14, 9$). Energy conservation imposes that the sum of the energies of the landing sites is equal to $20$, although the sum of the obtained energies is $40$. The rescaling parameter $\phi_\mu$ will be thus $40/20=2$. (d) We rescale all the outcomes by the factor $\phi_\mu$ and project to a new staircase. The effect of the rescaling is that, in this particular realisation of there cascading process, any value $E' $ such that $\phi_\mu E'>20$ will be forbidden --grey region in the picture. A cascade would iterate the described process until all created balls reach state $1$.}
\label{fig:Rescaling}
\end{figure}

In the sequel we derive the statistics of visits to the states of our system when our cascade observes an {\em energy conservation} constraint. Remarkably, we see that the histogram of visits to each state along the whole cascading process follows a power-law of exponent $-2$, regardless the multiplication parameter of the avalanche. The strategy followed here is based on the {\em partition through renormalization} and works basically as follows: out of an interval $[0,1]$, a partition is performed by throwing $\mu$ random numbers between $0$ and $1$ and then renormalizing them such that their sum is $1$ as in figure (\ref{fig:Rescaling}a,b) \cite{Poisson:1993}. Our strategy can be seen as a particular choice of Dirichlet partitioning of the interval \cite{Huillet:2005}. This strategy is different from the random selection of breaking points of the interval, described, e.g., in \cite{Feller:1959, Krapivsky:1994}. In the appendix A we provide a complete proof of our result.

To study SSR cascades with a superimposed conservation law, let us assume, with any loss of generality, that the states $1,2, \cdots,N$ are associated with energy levels 
\[
E_1,E_2, \cdots ,E_N\quad.
\]
Energy conservation imposes the following constraint: If a particle with energy $E$ and splits into $\mu$ particles $i_1, i_2, \cdots i_\mu$, with respective energies $E_{i_1},E_{i_2},\cdots ,E_{i_\mu}$, then: 
\begin{equation}
	\sum_{k=1}^\mu E_{i_k}=E\quad.
\label{eq:EnergiCons}
\end{equation}
We are interested in the energy spectrum, i.e. number of observations of particles at a particular energy level at any point of the cascading process, $n(E)$. 

\begin{figure}[ht!]
\includegraphics[width= 8.0cm]{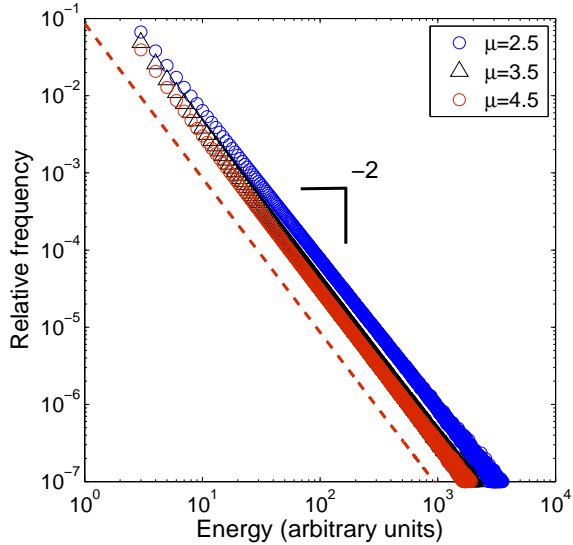}
\caption{Histograms of SSR cascades with energy conservation at each multiplication event, for $\mu=2.5$ (blue dots), $\mu=3.5$ (red triangles) and 
$\mu=4.5$ (black diamonds). As predicted only exponents with a value of $2$ occur. The dashed line shows the slope a perfect power-law with exponent $\alpha=2$ for comparison.
Curves were fitted using the matlab  \texttt{rplfit} package using likelihood estimations for all exponents \cite{Hanel:2016}. 
}
\label{fig:AvalanchesECons}
\end{figure}

The first task is to impose the energy conservation constraint given in equation (\ref{eq:EnergiCons}) in the schema of transition probabilities of the SSR cascade. To compute it we use a rescaling technique and we will assume that the energy spectrum is continuous. The rescaling technique is outlined in figure (\ref{fig:Rescaling}).  
Let us ignore energy conservation for the moment and 
define a continuous uniform random variable $u$ on the interval $[0,1]$. 
Let $u_1, u_2, \cdots ,u_\mu$ be $\mu$ independent realisations of $u$, see figure (\ref{fig:Rescaling}a). Let us suppose that we are at level $E$. From this sequence of random variables one can derive the target sites of the newly 
created particles in a SSR avalanche with multiplicative parameter $\mu$ as
\[
u_1\cdot E,u_2\cdot E, \cdots ,u_\mu \cdot E\quad.
\]
This is the continuous version of what we described in section \ref{Sec:Cascades}.
Now we define a new random variable, $\phi_\mu$, which is the sum of $\mu$ realisations of the random variable $u$:
\[
\phi_\mu=\sum_{k\leq \mu}u_k\quad.
\]
The sum of $\mu$ realisations of a random variable $u$ uniformly distributed on the interval $[0,1]$, $\phi_\mu$, follows the Irwin-Hall distribution, $f_\mu(\phi_\mu)$ \cite{Feller:1959}. This means that one can construct, for each $\mu$ realisations of the random variable $u$ a rescaled sequence, see figure (\ref{fig:Rescaling}b)
\[
\phi_\mu^{-1}u_1\cdot E, . . .,\phi_\mu^{-1}u_\mu \cdot E\quad,
\]
such that sum up to the total energy $E$, 
\begin{equation}
	\phi_\mu^{-1}\sum_{i\leq \mu}u_i\cdot E=E\quad.\nonumber
\end{equation}
Thus, by imposing energy conservation we actually expect the following sequence of rescaled energies $E_{i_1},E_{i_2},\cdots ,E_{i_\mu}$  for the emerging particles, where
\begin{equation}
	{E}_{i_k} =  \phi_\mu^{-1} (u_k \cdot E)\quad.\nonumber
\end{equation}
This rescaling approach assumes that the $\mu$ new particles behave independently. 
The crucial issue is to map this process into a cascade, see figure (\ref{fig:Rescaling}c,d).
To approach this problem, we first study the  expected number of particles that jump to a given state $E$ if a given value of $\phi_\mu$ occurs. We then average over all potential values of $\phi_\mu$. We assume that the expected number of particles from $E\to E'$, $n(E\to E')$ goes as $\sim \mu p(E'|E)$, as we did above. Taking into account the rescaling imposed by energy conservation, $p(E'|E)\sim \frac{\phi_\mu}{E}$, see figure (\ref{fig:Rescaling})-- on has that: 
\begin{equation}
	n(E\to E',\phi_\mu)=\left\{\begin{array}{cl}
	\mu\frac{\phi_\mu}{E}   & {\rm for}\;   \phi_\mu E'<E \\
	0                                   & {\rm for}\;   \phi_\mu E'\geq E\quad.
\end{array}
\right.
\label{eq:P(a_k|a_i)Energie}
\end{equation}
For a given value of $\phi_\mu$, that the expected number of particles that will visit state $E$ at some point of the cascade, $n(E,\phi_\mu)$ is:
\[
n(E,\phi_\mu)=\int_{\phi_\mu E}^\infty\mu\frac{\phi_\mu}{E'}n(E')dE'\quad,
\]
where $n(E')$ is the total number of particles that are expected to visit state $E'$ during the cascade. $n(E)$ will be obtained by averaging $n(E,\phi_\mu)$ over all potential values of $\phi_\mu$, distributed as the Irwin-Hall distribution, $f_\mu$:
\begin{equation}
n(E)=\int_0^\mu f_\mu(\phi_\mu)\left\{\int_{\phi_\mu E}^\infty\mu\frac{\phi_\mu}{E'}n(E')dE'\right\}d\phi_\mu\quad.
\label{eq:n(E)}
\end{equation}
Differentiating $n(E)$, one arrives at the following equation with displacement:
\[
\frac{dn}{dE}=-\mu\int_0^\mu \phi_\mu f_\mu(\phi_\mu)n(\phi_\mu E)d\phi_\mu\quad.
\]
Assuming that $n(E)\propto E^{-\alpha}$, we arrive at the following self-consistent equation for $\alpha$:
\begin{equation}
\alpha=\mu\int_0^\mu \phi_\mu^{1-\alpha} f_\mu(\phi_\mu)d\phi_\mu\quad.
\label{eq:alphamu}
\end{equation}
whose only solution, for large $\mu's$ converges to $\alpha=2$, leading to the general result.
\begin{equation}
	n(E)\propto E^{-2} \quad . 
\end{equation}
In the appendix A we provide a rigorous derivation of this result. In spite of the asymptotic nature of the proof given in the appendix A, numerical simulations show an excellent agreement with this theoretical prediction, even for $\mu$ small. In figure (\ref{fig:AvalanchesECons}) we show the frequency plots for $10^5$ avalanches with $\mu=2.5,3.5, 4.5$, and the convergence of the histograms to $E^{-2}$ can be perfectly appreciated. This result is the same that is obtained from Fermi's particle acceleration model to explain the spectrum of cosmic rays 
\cite{Fermi:1949, Longair:2008}. Here we derived it on the basis of  simple combinatorial reasonings of SSR processes.

%

\section{Discussion}

We have shown, both analytically and numerically, that SSR processes represent a general route to scaling, 
that is able to generate {\em any} scaling exponent. To obtain exponents larger than $1$, which was the main target of this paper, 
we introduced the concept of SSR cascades, a simple multiplicative process which combines SSR- and cascading processes. 
Our results also add a new way of the interpretation of scaling laws observed in multiplicative processes, and avalanche- or cascading processes. 
The quality of the SSR view rests in its extreme simplicity, intuitive nature and its generality. In addition, SSR cascades can be mapped to physical process of successive disintegration. By applying the approach to a physical cascading process of particle cascades from cosmic rays it is sufficient to impose energy conservation in 
the cascading process. In doing so, we recover the classic result of Fermi for the energy spectrum of cosmic rays. 
More generally, we have shown that energy conservation (or any other similar constraint) 
leads to a universal scaling exponent of $2$, regardless of the details of the microscopic cascading process.  
In the case of no constraint and no cascading present the ubiquitous Zipf's law is obtained, with its exponent $1$.  
Our findings imply that the presence of simple history-dependence imposed by SSR processes 
deforms statistics of in a highly non-intuitive way and leads naturally to power laws, and at the same 
time explains why exponents $1$ and $2$ are special. 
We believe that our results  might be useful to understand scaling in 
problems of statistical inference whenever observation biases  exist, 
for problems of fragmentation and cascading, including the meteorite energy spectrum, particle cascades, 
and for multiplicative directed diffusion on networks such as e.g. rumour spreading, or the spreading of viral loads in populations.

\newpage

\appendix
\section{Derivation of exponent $-2$ for cascades with energy conservation }

\begin{figure}
\includegraphics[width= 8.0cm]{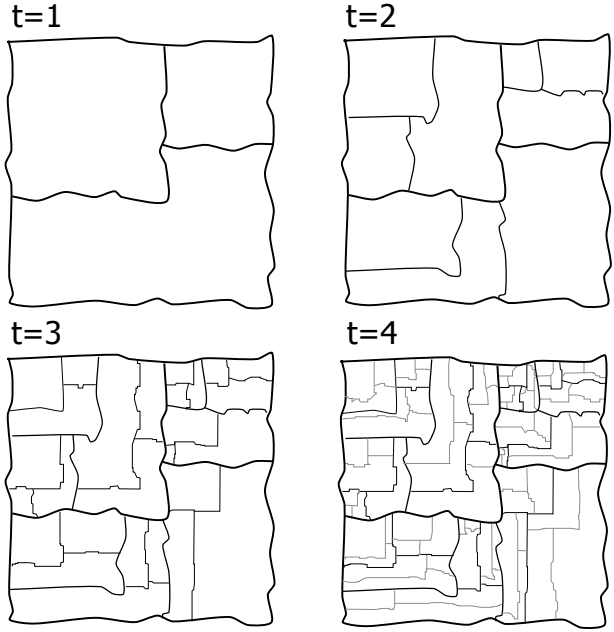}
\caption{SSR cascades with energy conservation can be seen as follows: Consider a polygon with a given area (e.g., $1$), $t=0$. At $t=1$ throw three random numbers between $0$ and $1$, $x_1,x_2,x_3$ and then rescale them $x'_i=x_i/(x_1+x_2+x_3)$. Now embed three polygons of arbitrary shape with area $x'_1,x'_2,x'_3$, respectively, and put them inside the first polygon. At $t=2$ we repeat the operation for each of the subpolygons created and we iterate the process for $t\gg 1$.Notice that the 2D polygon is just a representation and that our result does not depend on the dimension of the partitioned object.}
\label{fig:Partition_Square}
\end{figure}

We derive the main result of section IIB. The strategy followed is summarised in figure (\ref{fig:Rescaling}). An alternative view is given in figure (\ref{fig:Partition_Square}) in this appendix.

Let $u$ be a random variable whose probability density is uniform in the interval $[0,1]$. Let $u_1, . . .,u_\mu$ be a sequence of independent drawings of the random variable $u$ and $\phi(\mu)$ a random variable defined over the interval $[0,\mu]$ as:
\begin{equation}
\phi_\mu=\sum_{k\leq \mu}u_k\quad.
\label{eq:Irwin_hallRV}
\end{equation}
The probability density that governs the random variable $\phi_\mu$ is the Irwin-Hall distribution. Now we go to equation (\ref{eq:n(E)}),
\begin{equation}
n(E)=\int_0^\mu f_\mu(\phi_\mu)\left\{\int_{\phi_\mu E}^\infty\mu\frac{\phi_\mu}{E'}n(E')dE'\right\}d\phi_\mu\quad.\nonumber
\end{equation}
Differentiating,
\begin{eqnarray}
\frac{dn}{dE}&=&\frac{d}{dE}\int_0^\mu f_\mu(\phi_\mu)\left\{\int_{\phi_\mu E}^\infty\mu\frac{\phi_\mu}{E'}n(E')dE'\right\}d\phi_\mu\nonumber\\
&=&\int_0^\mu f_\mu(\phi_\mu)\frac{d}{dE}\left\{\int_{\phi_\mu E}^\infty\mu\frac{\phi_\mu}{E'}n(E')dE'\right\}d\phi_\mu\nonumber\\
&=&-\mu\int_0^\mu \phi_\mu f_\mu(\phi_\mu)n(\phi_\mu E)d\phi_\mu\quad,\nonumber
\end{eqnarray}
one arrives at the following equation with displacement:
\[
\frac{dn}{dE}=-\mu\int_0^\mu \phi_\mu f_\mu(\phi_\mu)n(\phi_\mu E)d\phi_\mu\quad.
\]
Assuming that $n(E)\propto E^{-\alpha}$, we arrive at the following self-consistent equation for $\alpha$:
\begin{equation}
\alpha=\mu\int_0^\mu \phi_\mu^{1-\alpha} f_\mu(\phi_\mu)d\phi_\mu\quad.\nonumber
\end{equation}
This is equation (\ref{eq:alphamu}) and that is what we have to solve. The following theorem states that the solution in the limit of large $\mu$'s is $\alpha=2$, independent of $\mu$. To demonstrate that we need to proof $5$ lemmas. After the demonstration, we approach the solution $\alpha\to 2$ using a mean field approach. Finally, we report a side observation concerning the behaviour of the average value of a random variable following the Irwin-hall distribution.
$\\$

\noindent
{\bf Theorem}: {\em The only $\alpha$ satisfying the following equation:}
\begin{equation}
\lim_{\mu\to\infty}\mu\int_0^\mu \phi_\mu^{1-\alpha} f_\mu(\phi_\mu)d\phi_\mu=\alpha\quad.
\label{eq:theorem}
\end{equation}
{\em is $\alpha=2$.}

To prove this theorem, we first observe that a random variable following the Irwin-Hall distribution converges to a random variable following a normal distribution with average $\frac{\mu}{2}$ and standard deviation $\sqrt{\frac{\mu}{12}}$. 
$\\$

\noindent
{\bf Lemma 1}. {\em Let $\phi_\mu$ be a random variable following the Irwin-Hall distribution. Let $Y$ be a random variable following a normal distribution centred at $0$ and with standard deviation $1$. Then, the following limit holds:}
\[
\phi_\mu\to \frac{\mu}{2}+\sqrt{\frac{\mu}{12}}Y(0,1)\quad.
\]
{\em in probability.}
 
\noindent
{\bf Proof}.
The average value of a uniformly distributed random variable $u$ is $\mathbb{E}(u)=\frac{1}{2}$ and standard deviation $\sigma=\sqrt{\frac{1}{12}}$. 
By the central limit theorem, one has that, for an i.i.d. sequence of random variables $u_1, . . . ,u_n$:
\[
\frac{\sum_{i\leq \mu}u_i-\frac{\mu}{2}}{\sqrt{\frac{\mu}{12}}}\to Y\quad,
\]
being $Y$ a random variable following a normal distribution centred at $0$ and with standard deviation $1$. Therefore, by realising that $\phi_\mu$ is actually a sum of $\mu$ i.i.d random variables $u$, one has:
\[
\phi_\mu\to \frac{\mu}{2}+\sqrt{\frac{\mu}{12}}Y(0,1)\quad,
\]
as we wanted to demonstrate.$\square$
$\\$

This implies that the Irwin-Hall distribution $f_\mu$ can be fairly approached by a normal distribution with mean $\frac{\mu}{2}$ and standard deviation $\sqrt{\frac{\mu}{12}}$, $\Phi(x)$. However, one must be careful with this approach: It can lead the integral that we want to solve, $\int_0^\mu \phi_\mu^{1-\alpha} f_\mu(\phi_\mu)d\phi_\mu$, to a singularity at $0$ which is inexistent in the Irwin-Hall distribution. Therefore, for the interval $[0,1)$ we will maintain the original form of the distribution. In the following lemma we demonstrate that this has no impact in the limit of large $\mu$'s.
$\\$

\noindent
{\bf Lemma 2}: {\em Let $\Phi_\mu$ be a normal distribution with mean at $\frac{\mu}{2}$ and standard deviation $\sigma_\mu=\sqrt{\frac{\mu}{12}}$. Then, $(\forall\epsilon >0)(\exists N):(\forall \mu>N)$}
\[
\left|\int_0^\mu \phi_\mu^{1-\alpha} f_\mu(\phi_\mu)d\phi_\mu-\int_1^\mu \phi_\mu^{1-\alpha} \Phi(\phi_\mu)d\phi_\mu\right|<\epsilon\quad.
\]

\noindent
{\bf Proof}: From lemma 1 we know that $(\forall\epsilon ' >0)(\exists N):(\forall \mu>N)$
\[
\left|\int_1^\mu \phi_\mu^{1-\alpha} f_\mu(\phi_\mu)d\phi_\mu-\int_1^\mu \phi_\mu^{1-\alpha} \Phi_\mu(\phi_\mu)d\phi_\mu\right|<\epsilon '\quad.
\]
Now we use the fact that the Irwin-Hall distribution can be defined per intervals using different polynomials. In the case of the interval $[0,1)$, the polynomial reads: 
\begin{eqnarray}
f_\mu(\phi_\mu)=\frac{1}{(\mu-1)!}\phi_\mu^{\mu-1};\quad\phi_\mu\in [0,1)\quad.
\label{eq:[0,1)}
\end{eqnarray}
Computing directly the integral, one has:
\begin{eqnarray}
\int_0^\mu \phi_\mu^{1-\alpha} f_\mu(\phi_\mu)d\phi_\mu&=&\int_0^1\phi_\mu^{1-\alpha} f_\mu(\phi_\mu)d\phi_\mu+\nonumber\\
&&+\int_1^\mu \phi_\mu^{1-\alpha} f_\mu(\phi_\mu)d\phi_\mu\quad,\nonumber
\end{eqnarray}
where the first integral, according to equation (\ref{eq:[0,1)}) leads to:
\begin{eqnarray}
\int_0^1\phi_\mu^{1-\alpha} f_\mu(\phi_\mu)d\phi_\mu&=&\frac{1}{(\mu-1)!}\int_0^1\phi_\mu^{\mu-\alpha}d\phi_\mu\nonumber\\
&=&\frac{1}{(\mu-1)!(\mu-\alpha-1)}\quad.\nonumber
\end{eqnarray}
Now take $\delta\in (0,1)$ and define $\epsilon(\mu, \delta)$ as:
\[
\epsilon(\mu)\equiv\frac{1+\delta}{(\mu-1)!(\mu-\alpha-1)}
\]
From lemma 1, $(\forall\epsilon ' >0)(\exists N):(\forall \mu>N)$ we can define the following bound:
\[
\left|\int_0^\mu \phi_\mu^{1-\alpha} f_\mu(\phi_\mu)d\phi_\mu-\int_1^\mu \phi_\mu^{1-\alpha} \Phi_\mu(\phi_\mu)d\phi_\mu\right|<\epsilon'+\epsilon(\mu,\delta) \;.
\]
Finally, we observe that
\[
\lim_{\mu\to\infty}\epsilon(\mu,\delta)=0\quad,
\]
which demonstrates the lemma.$\square$
$\\$

\noindent
{\bf Lemma 3}: {\em The function of $G(\alpha)$ defined by the integral}
\[
G(\alpha)=\int_1^\mu \phi_\mu^{1-\alpha} \Phi(\phi_\mu)d\phi_\mu\quad,
\]
{\em is strictly decreasing}.

\noindent
{\bf Proof}: It is enough to compute the derivative:
\begin{eqnarray}
\frac{d}{d\alpha}\int_1^\mu \phi_\mu^{1-\alpha} \Phi(\phi_\mu)d\phi_\mu&=&\int_1^\mu \left(\frac{d}{d\alpha}\phi_\mu^{1-\alpha}\right) \Phi(\phi_\mu)d\phi_\mu\nonumber\\
&=&-\int_1^\mu \phi_\mu^{1-\alpha}\log\phi_\mu \Phi(\phi_\mu)d\phi_\mu\nonumber\\
&<& 0\quad,\nonumber
\end{eqnarray}
since the term inside the integral, $ \phi_\mu^{1-\alpha}\log\phi_\mu \Phi(\phi_\mu)$, is strictly positive in the interval $(1,\mu)$. $\square$
$\\$

Now take a monotonously increasing function that grows slower than the standard deviation $\sigma_\mu=\sqrt{\frac{\mu}{12}}$, $\varphi(\mu)$. For convenience, we define define it as:
\begin{equation}
\varphi(\mu)\equiv\left(\frac{\mu}{12}\right)^{\frac{1}{4}}\quad.
\label{eq:varphi}
\end{equation}
Clearly:
\begin{equation}
\lim_{\mu\to\infty}\int_{\frac{\mu}{2}-\sigma_\mu\varphi(\mu)}^{\frac{\mu}{2}+\sigma_\mu\varphi(\mu)} \Phi(\phi_\mu)d\phi_\mu=1\quad.
\label{eq:normalized}
\end{equation}
Now suppose that $\alpha=2$. Then, thanks to Lemma 2, one has that
$(\forall\epsilon >0)(\exists N):(\forall \mu>N)$
\[
\left|\int_0^\mu \frac{f_\mu(\phi_\mu)}{\phi_\mu} d\phi_\mu-\int_1^\mu \frac{\Phi_\mu(\phi_\mu)}{\phi_\mu}d\phi_\mu\right|<\epsilon\quad.
\]
From this we derive the third lemma of our demonstration:
$\\$

\noindent
{\bf Lemma 4}: $(\forall\epsilon >0)(\exists N):(\forall \mu>N),$
\[
\left|\int_1^\mu \frac{\Phi(\phi_\mu)}{\phi_\mu }d\phi_\mu-\int_{\frac{\mu}{2}-\sigma_\mu\varphi(\mu)}^{\frac{\mu}{2}+\sigma_\mu\varphi(\mu)} \frac{\Phi(\phi_\mu)}{\phi_\mu }d\phi_\mu\right|<\epsilon\quad.
\]
\noindent
{\bf Proof}:
We need to compute the parts that fall outside the integration limits and see that their contribution vanishes. First, we see that:
\begin{eqnarray}
\int_{\frac{\mu}{2}+\sigma_\mu\varphi(\mu)}^\mu  \frac{\Phi(\phi_\mu)}{\phi_\mu }d\phi_\mu &<& \left(\frac{\mu}{2}+\sigma_\mu\varphi(\mu)\right)e^{-\varphi^2(\mu)-{\cal O}(\log\mu)}\nonumber\\
&<& \left(\frac{\mu}{2}+\sigma_\mu\varphi(\mu)\right)e^{-\sqrt{\mu}}\quad.\nonumber
\end{eqnarray}
Analogously, 
\begin{eqnarray}
\int^{\frac{\mu}{2}-\sigma_\mu\varphi(\mu)}_1  \frac{\Phi(\phi_\mu)}{\phi_\mu }d\phi_\mu< \left(\frac{\mu}{2}-\sigma_\mu\varphi(\mu)\right)e^{-\sqrt{\mu}}\quad.\nonumber
\end{eqnarray}
Now we define:
\begin{eqnarray}
\epsilon_1(\mu)&=&\left(\frac{\mu}{2}+\sigma_\mu\varphi(\mu)\right)e^{-\sqrt{\mu}}\quad,\nonumber\\
\epsilon_2(\mu)&=&\left(\frac{\mu}{2}-\sigma_\mu\varphi(\mu)\right)e^{-\sqrt{\mu}}\quad,\nonumber
\end{eqnarray}
which leads to:
\[
\left|\int_1^\mu \frac{\Phi(\phi_\mu)}{\phi_\mu }d\phi_\mu-\int_{\frac{\mu}{2}-\sigma_\mu\varphi(\mu)}^{\frac{\mu}{2}+\sigma_\mu\varphi(\mu)} \frac{\Phi(\phi_\mu)}{\phi_\mu }d\phi_\mu\right|<\epsilon_1(\mu)+\epsilon_2(\mu)\;,
\]
demonstrating the lemma.$\square$
$\\$

\noindent
{\bf Corollary of Lemma 4}:  $(\forall\epsilon >0)(\exists N):(\forall \mu>N),$
\[
\left|\int_0^\mu \phi_\mu^{1-\alpha} f_\mu(\phi_\mu)d\phi_\mu-\int_{\frac{\mu}{2}-\sigma_\mu\varphi(\mu)}^{\frac{\mu}{2}+\sigma_\mu\varphi(\mu)} \frac{\Phi(\phi_\mu)}{\phi_\mu }d\phi_\mu\right|<\epsilon\quad.
\]

\noindent
{\bf Proof}:
By direct application of lemmas 2 and 4.$\square$
$\\$

Now we define the following functions of the limits of the integral $\int_{\frac{\mu}{2}-\sigma_\mu\varphi(\mu)}^{\frac{\mu}{2}+\sigma_\mu\varphi(\mu)}\dots$:
\begin{eqnarray}
r_1(\mu)\equiv\left(\frac{\mu}{2}+\sigma_\mu\varphi(\mu)\right)^{-1}=\frac{2}{\mu}-\frac{2\sigma_\mu\varphi_\mu}{\mu\left(\frac{\mu}{2}+\sigma_\mu\varphi(\mu)\right)}\;,\nonumber\\
r_2(\mu)\equiv\left(\frac{\mu}{2}-\sigma_\mu\varphi(\mu)\right)^{-1}=\frac{2}{\mu}+\frac{2\sigma_\mu\varphi_\mu}{\mu\left(\frac{\mu}{2}-\sigma_\mu\varphi(\mu)\right)}\;.
\nonumber
\end{eqnarray}
Clearly,
\begin{equation}
r_{1,2}(\mu)\sim \frac{2}{\mu}+{\cal O}\left(\mu^{-\frac{5}{4}}\right)\quad,
\label{eq:r12}
\end{equation}
where the subscript $_{1,2}$ means that both functions satisfy the property.
$\\$

\noindent
{\bf Lemma 5}. $(\forall\epsilon >0)(\exists N):(\forall \mu>N)$
\begin{eqnarray}
\left|r_{1,2}(\mu)\cdot\mu\int_{\frac{\mu}{2}-\sigma_\mu\varphi(\mu)}^{\frac{\mu}{2}+\sigma_\mu\varphi(\mu)} \right.&&\Phi(\phi_\mu)d\phi_\mu-\nonumber\\
&&-\left.\mu\int_{\frac{\mu}{2}-\sigma_\mu\varphi(\mu)}^{\frac{\mu}{2}+\sigma_\mu\varphi(\mu)} \frac{\Phi(\phi_\mu)}{\phi_\mu }d\phi_\mu\right|<\epsilon\;.\nonumber
\end{eqnarray}

\noindent
{\bf Proof}. 
We first observe that, by substituting $\phi^{-1}_\mu$ by the integration limits, we have the following chain of inequalities, in terms of the above defined functions $r_{1,2}(\mu)$:
\begin{eqnarray}
r_{2}(\mu)&\cdot&\int_{\frac{\mu}{2}-\sigma_\mu\varphi(\mu)}^{\frac{\mu}{2}+\sigma_\mu\varphi(\mu)} \Phi(\phi_\mu)d\phi_\mu<\nonumber\\
 &<&\int_{\frac{\mu}{2}-\sigma_\mu\varphi(\mu)}^{\frac{\mu}{2}+\sigma_\mu\varphi(\mu)} \frac{\Phi(\phi_\mu)}{\phi_\mu }d\phi_\mu\nonumber\\
&<&r_{1}(\mu)\cdot\int_{\frac{\mu}{2}-\sigma_\mu\varphi(\mu)}^{\frac{\mu}{2}+\sigma_\mu\varphi(\mu)} \Phi(\phi_\mu)d\phi_\mu\quad.\nonumber
\end{eqnarray}
Therefore, is enough to demonstrate that $(\forall\epsilon' >0)(\exists N):(\forall \mu>N)$
\begin{equation}
\left|\mu r_1(\mu)-\mu r_2(\mu)\right |<\epsilon'\quad.\nonumber
\end{equation}
This can be proven directly from equation (\ref{eq:r12}), leading to:
\begin{equation}
\left|\mu r_1(\mu)-\mu r_2(\mu)\right|\sim {\cal O}\left(\mu^{-\frac{1}{4}}\right)\quad,\nonumber
\end{equation}
which demonstrates the lemma.$\square$
$\\$

Collecting lemmas 1,2,4, and 5, we have demonstrated that, under the assumption that $\alpha=2$,
\[
\mu r_1(\mu)\to \mu\int_0^\mu \phi_\mu^{1-\alpha} f_\mu(\phi_\mu)d\phi_\mu\quad.
\]
From equation (\ref{eq:alphamu}), the only remaining issue to demonstrate the consistency of our hypothesis is that indeed $\lim_{\mu_\to\infty}\mu r_1(\mu)=2$. It is not difficult to check, from equation (\ref{eq:r12}), that:
\[
\lim_{\mu\to \infty} \mu r_1(\mu) =\lim_{\mu\to \infty} \left[ 2+{\cal O}\left(\mu^{-\frac{1}{4}}\right)\right]=2 \quad.
\] 
So far we have demonstrated that the solution $\alpha=2$ is consistent with the statement of the theorem. Now it remains to demonstrate that this is the only solution.
To see that, we observe that thanks to lemma 3, we know that the function $\mu G(\alpha)$ is decreasing on $\alpha$. In addition, we have proven that the statement of the theorem is consistent for $\alpha=2$. Therefore, if $\alpha=2+\beta$, with $\beta>0$, then:
\[
\mu\int_0^\mu \phi_\mu^{1-\alpha} f_\mu(\phi_\mu)d\phi_\mu<2+\beta\quad,
\]
which contradicts the statement of the theorem. The same happens if one imposes if $\alpha=2-\beta$, with $\beta>0$, since one gets:
\[
\mu\int_0^\mu \phi_\mu^{1-\alpha} f_\mu(\phi_\mu)d\phi_\mu>2-\beta\quad,
\]
which is, again inconsistent, thereby proving the lemma.$\square$
$\\$

By direct application of equation (\ref{eq:alphamu}), Lemma 5 puts the last piece to demonstrate the theorem. $\blacksquare$
$\\$

{\em Mean-field approach}.- In a less rigorous way, we observe that we can approach the solution as follows: We know that the expected value of  a random variable $\phi_\mu$ following the Irwin-Hall distribution $f_\mu$ is:
\[
\mathbb{E}(\phi_\mu)=\frac{\mu}{2}\quad.
\]
Now assume that $\phi_\mu\approx \frac{\mu}{2}$. This implies that, in the integral of the statement of the theorem, equation (\ref{eq:theorem}), we replace $f_\mu(\phi_\mu)$ by $\delta\left(\phi_\mu-\frac{\mu}{2}\right)$, where $\delta$ is the Dirac $\delta$ function:
\[
\mu\int_0^\mu \phi_\mu^{1-\alpha} f_\mu(\phi_\mu)d\phi_\mu\approx\mu\int_0^\mu \phi_\mu^{1-\alpha} \delta\left(\phi_\mu-\frac{\mu}{2}\right)d\phi_\mu\quad.
\]
Solving the integral, and thanks to equation (\ref{eq:alphamu}), we obtain the following relation:
\[
\frac{\alpha}{\mu}=\left(\frac{\mu}{2}\right)^{1-\alpha}\quad,
\]
whose only solution is $\alpha=2$.
$\\$

We end by observing that the theorem that we demonstrated has a curious consequence: Let $\phi_\mu$ be a random variable following the Irwin-Hall distribution. We observe that, if $\alpha=2$, then:
\begin{eqnarray}
\frac{\alpha}{\mu}&=&\int_0^\mu \phi_\mu^{1-\alpha} f_\mu(\phi_\mu)d\phi_\mu\nonumber\\
&=&\int_0^\mu \frac{f_\mu(\phi_\mu)}{\phi_\mu }d\phi_\mu\nonumber\\
&=&\mathbb{E}\left(\frac{1}{\phi_\mu}\right)\nonumber\\
&\to&\frac{2}{\mu}\quad.\nonumber
\end{eqnarray}
We know that $\mathbb{E}(\phi_\mu)=\frac{\mu}{2}$. Therefore, a direct consequence of the theorem is that:
\[
\mathbb{E}\left(\frac{1}{\phi_\mu}\right)\to\frac{1}{\mathbb{E}(\phi_\mu)}\quad.
\]

\section*{Additional information}
$\\$
{\bf Acknowledgments}
$\\$
This work was supported by the Austrian Science Fund FWF under the P29232 and P 29252 projects. We acknowledge an anonymous reviewer for the constructive comments on our first version of the manuscript.
$\\$
{\bf Author contributions}
$\\$
B C-M, R H and S T designed, performed the research and wrote the manuscript.
$\\$
{\bf Competing interests}
$\\$
The authors declare no competing financial interest.

\end{document}